\documentclass[twocolumn,10pt,aip,jcp,amsmath,amssymb]{revtex4-1}
\usepackage{pdfpages}
\usepackage[utf8]{inputenc}
\usepackage{amsmath}
\usepackage{graphicx}   % include figure files
\usepackage{dcolumn}    % align numerical table columns on the decimal point
\usepackage{bm}         % bold math symbols
\usepackage{hyperref}
\usepackage{color}
\usepackage{array}
\usepackage{setspace}
\newcolumntype{L}[1]{>{\raggedright\let\newline\\\arraybackslash\hspace{0pt}}m{#1}}

\newcommand{\wt}{\widetilde}

\newcommand{\jcp}{J.\ Chem.\ Phys.\ }

\newcommand{\jpca}{J.\ Phys.\ Chem.\ A }

\newcommand{\tr}{ {\rm Tr} }
\newcommand{\etf}{e^{-i \hat H t/\hbar}}
\newcommand{\etb}{e^{i \hat H t/\hbar}}
\newcommand{\no}{\nonumber}
\newcommand{\bq}{ {\bf q} }

\newcommand{\eqn}[1]{Eq.~(\ref{#1})}
\newcommand{\Eqn}[1]{Equation~(\ref{#1})}

\begin{document}

\title{Relation of centroid molecular dynamics and ring-polymer molecular dynamics to exact quantum dynamics} 
%\author{Timothy J.~H.~Hele, Michael J.~Willatt, Andrea~Muolo\footnote{Current address: Lab.\ f\" ur Physikalische Chemie, ETH Z\" urich, CH-8093 Z\" urich, Switzerland.} and Stuart C.~Althorpe\footnote{Corresponding author: sca10@cam.ac.uk}}
%\affiliation{Department of Chemistry, University of Cambridge, Lensfield Road, Cambridge, CB2 1EW, UK.}
\author{Timothy J.~H.~Hele} 
\noaffiliation
\author{Michael J.~Willatt}
\noaffiliation
\author{Andrea~Muolo}
\email[Current address: Lab. für Physikalische Chemie, ETH Zürich, CH-8093 Zürich, Switzerland]{}
\noaffiliation
\author{Stuart C.~Althorpe}
\email[Corresponding author: ]{sca10@cam.ac.uk}
\noaffiliation
\affiliation{\mbox{Department of Chemistry, University of Cambridge, Lensfield Road, Cambridge, CB2 1EW, UK.}}
\date{\today}

\begin{abstract}
We recently  obtained a quantum-Boltzmann-conserving classical dynamics by
making a single change to the derivation of the `Classical Wigner'
approximation. Here, we show that the further approximation of this `Matsubara
dynamics' gives rise to two popular heuristic methods for treating quantum
Boltzmann time-correlation functions: centroid molecular dynamics (CMD) and
ring-polymer molecular dynamics (RPMD). We show that CMD is a mean-field
approximation to Matsubara dynamics, obtained by discarding (classical)
fluctuations around the centroid, and that RPMD is the result of discarding a
term in the Matsubara Liouvillian which shifts the frequencies of these
fluctuations.  These findings are consistent with previous numerical results,
and give explicit formulae for the terms that CMD and RPMD leave out.
\textit{Copyright (2015) American Institute of Physics. This article may be
    downloaded for personal use only. Any other use requires prior permission of
the author and the American Institute of Physics.  The following article
appeared in the Journal of Chemical Physics, 142, 191101 (2015) and may be found
at http://dx.doi.org/10.1063/1.4921234}
\end{abstract}

\maketitle 

\section{Introduction}

Quantum Boltzmann time-correlation functions play a central role in chemical physics, and are (usually) impossible to calculate exactly. One promising approach is to treat  the statistics quantally and the dynamics classically. The standard way to do this is to use the linearized semi-classical initial value representation (LSC-IVR or `classical Wigner approximation'), \cite{billrev} but this has the drawback of not satisfying detailed balance. Recently,\cite{mats} however, we found that a single change to the LSC-IVR derivation gives a classical dynamics which does satisfy detailed balance. This modified version of LSC-IVR is called `Matsubara dynamics'.

We can summarise Matsubara dynamics as follows: At initial time, the quantum statistics gives rise to delocalized distributions in position which are smooth functions of imaginary time. If we constrain the LSC-IVR dynamics to conserve this smoothness (by including only the smooth `Matsubara' modes---see Sec.~II) we find that it satisfies detailed balance, and gives  better agreement than LSC-IVR with the exact quantum result.\cite{mats} We suspect (but have not yet proved) that Matsubara dynamics reproduces the time-dependence of the exact Kubo-transformed time-correlation function up to order $\hbar^0$, and is thus the correct theory for describing  quantum statistics and classical dynamics.

Matsubara dynamics suffers from the sign problem and is hence impractical, 
but the findings just described suggest that it should be the starting point from which to make further approximations if one wishes to devise practical methods that combine quantum statistics with classical dynamics. Numerical tests in ref.~\onlinecite{mats} (see also Fig.~1) showed that the popular centroid molecular dynamics\cite{cmd1,cmd2} (CMD) and ring-polymer molecular dynamics\cite{rpmd1,rpmd2} (RPMD) methods appear to be two such approximations. Here we confirm this, by deriving the terms that CMD and RPMD leave out from the Matsubara dynamics. \cite{multi}

\section{Summary of Matsubara dynamics}
Matsubara dynamics approximates the quantum Kubo-transformed time-correlation function\cite{noz} 
\begin{align}
 C_{AB}(t) = {1\over\beta}\int_0^\beta d\lambda\,\tr\left[e^{-\lambda {\hat H}}\hat A e^{-(\beta-\lambda) {\hat H}}\etb \hat B \etf \right] \label{kubo}
\end{align}
\vspace{-0.3cm}
by 
\begin{align}
C_{ AB}^{\rm Mats}(t)=\lim_{M\to\infty} C_{ AB}^{[M]}(t)
 \end{align}
where
\begin{align}
 C_{ AB}^{[M]}(t) ={\alpha_M\over 2\pi\hbar}\int d{\bf \widetilde P}& \int d{\bf \widetilde Q}\ A({\bf \widetilde Q})e^{-\beta[{\wt H}_M({\bf \wt P},{\bf \wt Q})-i\theta_M({\bf \widetilde P},{\bf \widetilde Q})]}\no\\
&\times e^{{\cal L}_{M}t}B({\bf \widetilde Q})\label{mattsc}
\end{align}
and $\alpha_M=\hbar^{(1-M)}\left[(M-1)/ 2\right]!^2$.
The position coordinates ${\bf \widetilde Q}\equiv\{\wt Q_n\}$, with
$n=-(M-1)/2,\dots,(M-1)/2$, are the $M$ Matsubara  modes, which describe closed
paths $q(\tau)$ that are smooth functions of the imaginary time $\tau$
$(=0\to\beta\hbar)$, where $\wt Q_0$ is the centroid coordinate (see the Appendix);
$\int\!d{\bf\wt Q}\equiv \prod_n\int_{-\infty}^\infty\!d\wt Q_n$, and ${\bf
\widetilde P}$ are similarly defined for momentum. The functions $A({\bf
\widetilde Q})$ and $B({\bf \widetilde Q})$ are obtained from the operators
${\hat A}$ and ${\hat B}$ (see the Appendix), such that
${\hat A}={\hat B}={\hat q}$ gives $A({\bf \widetilde Q})=B({\bf \widetilde Q})=\wt Q_0$. \cite{alg}
The propagator $e^{{\cal L}_{M}t}$ contains the  Matsubara Liouvillian 
\begin{align}
{\cal L}_{M}=\sum_{n=-(M-1)/2}^{(M-1)/2}{{\widetilde P}_n\over m}{\partial \over \partial {\widetilde Q}_n}
  - {\partial {\widetilde U}_M({\bf \widetilde Q})\over\partial {\widetilde Q}_n}{\partial \over\partial {\widetilde P}_n}\label{matliou}
\end{align}
in which the potential energy ${\widetilde U}_M({\bf \widetilde Q})$ is  given
in the Appendix.  The quantum Boltzmann distribution in \eqn{mattsc} is complex,
and contains the Matsubara Hamiltonian
\begin{align}
{\wt H}_M({\bf \wt P},{\bf \wt Q})={{\bf \widetilde P}^2\over 2m}+{\widetilde U}_M({\bf \widetilde Q})
\end{align}
 and the phase
\begin{align}
\theta_M({\bf \widetilde P},{\bf \widetilde Q})=\sum_{n=-(M-1)/2}^{(M-1)/2}{\widetilde P}_n{\widetilde\omega}_n{\widetilde Q}_{-n}\label{thet}
\end{align}
where $\wt \omega_n$ are the Matsubara frequencies $\wt \omega_n=2\pi n/\beta\hbar$.
Matsubara dynamics is inherently classical (meaning that terms ${\cal O}(\hbar^2)$ disappear from the quantum Liouvillian on decoupling the Matsubara modes, leaving ${\cal L}_{M}$), and conserves the Hamiltonian ${\wt H}_M({\bf \wt P},{\bf \wt Q})$ {\em and} the phase $\theta_M({\bf \widetilde P},{\bf \widetilde Q})$, and thus satisfies detailed balance. 

Clearly \eqn{mattsc}  suffers from the sign problem because of the phase $\theta_M({\bf \widetilde P},{\bf \widetilde Q})$. Let us make  the coordinate transformation ${\overline P}_n={\wt P}_n-im{\wt \omega}_n{\wt Q}_{-n}$. This gives
%\begin{align}
% C_{ AB}^{[M]}(t) ={\alpha_M\over 2\pi\hbar}\left[\prod_{n=-(M-1)/2}^{(M-1)/2}\int_{-\infty-im{\wt \omega}_n{\wt Q}_{-n}}^{\infty-im{\wt \omega}_n{\wt Q}_{-n}}  d{\overline P}_n\right]& \int d{\bf \widetilde Q}\ A({\bf \widetilde Q})e^{-\beta \wt R_M
% ({\bf \overline P},{\bf \wt Q})}e^{{\cal L}_{M}t}B({\bf \widetilde Q})\label{trans}
%\end{align}
\begin{align}
 C_{ AB}^{[M]}(t) =& {\alpha_M\over 2\pi\hbar}\left[\prod_{n=-(M-1)/2}^{(M-1)/2}\int_{-\infty-im{\wt \omega}_n{\wt Q}_{-n}}^{\infty-im{\wt \omega}_n{\wt Q}_{-n}}  
                     d{\overline P}_n\right] \no \\
                   & \times \int d{\bf \widetilde Q}\ A({\bf \widetilde Q})e^{-\beta \wt R_M ({\bf \overline P},{\bf \wt Q})}e^{{\cal L}_{M}t}B({\bf \widetilde Q})
\label{trans}
\end{align}
where 
 \begin{align}
 \wt R_M({\bf \overline P},{\bf \wt Q}) =& \left(\sum_{n=-(M-1)/2}^{(M-1)/2}{{\overline P}_n^2\over 2m}+{m\over 2}\wt\omega_n^2\wt Q_n^2\right)+\wt U_M({\bf \wt Q})
 \end{align}
is the `ring-polymer' Hamiltonian familiar from quantum statistics.\cite{smooth,chanwol,parri,freeman} \Eqn{trans} is simply \eqn{mattsc} in disguise, but at $t=0$, we can use a standard contour-integration trick\cite{suppl} to shift
${\overline P_{n}}$ onto the real axis, giving 
 \begin{align}
 C_{ AB}^{[M]}(0) ={\alpha_M\over 2\pi\hbar}\int d{\bf \overline P}& \int d{\bf \widetilde Q}\ A({\bf \widetilde Q})B({\bf \widetilde Q})e^{-\beta \wt R_M
 ({\bf \overline P},{\bf \wt Q})}
\end{align}
which now contains the (real) ring-polymer  distribution,\cite{smooth} and hence no longer suffers from the sign problem.   Unfortunately, this trick does not work for $t>0$ (see Sec.~IV), so we are stuck with \eqn{mattsc}, which motivates us to find approximations to Matsubara dynamics.

\section{Centroid mean-field approximation}

This approximation can be made if $A({\bf \widetilde{Q}})$ is a function of just the centroid ${\wt Q}_0$ (or ${\wt P}_0$),\cite{alg} in which case we need only the
Matsubara dynamics of the centroid reduced density 
\begin{align}
b(\wt Q_0,\wt P_0,t)=\int d{\bf \widetilde P}'& \int d{\bf \widetilde Q}'\ e^{-\beta[{\wt H}_M({\bf \wt P},{\bf \wt Q})-i\theta_M({\bf \widetilde P},{\bf \widetilde Q})]}\no\\
&\times e^{{\cal L}_{M}t}B({\bf \widetilde Q})
\end{align}
where the primes denote integration over all modes except  ${\wt P}_0$ and ${\wt Q}_0$. Differentiation with respect to $t$,
 application of \eqn{matliou}, and integration by parts gives
\begin{align}
\dot b(\wt Q_0,\wt P_0,t)=\int d{\bf \widetilde P}'& \int d{\bf \widetilde Q}'\ e^{-\beta[{\wt H}_M({\bf \wt P},{\bf \wt Q})-i\theta_M({\bf \widetilde P},{\bf \widetilde Q})]}\no\\
&\times {\cal L}_0e^{{\cal L}_{M}t}B({\bf \widetilde Q})\label{dotb}
\end{align}
where
\begin{align}
{\cal L}_{0}={{\widetilde P}_0\over m}{\partial \over \partial {\widetilde Q}_0}
  - {\partial {\widetilde U}_M({\bf \widetilde Q})\over\partial {\widetilde Q}_0}{\partial \over\partial {\widetilde P}_0}
\end{align}
\begin{figure}[t]
    \begin{center}
        \centering
        \includegraphics[scale=0.70]{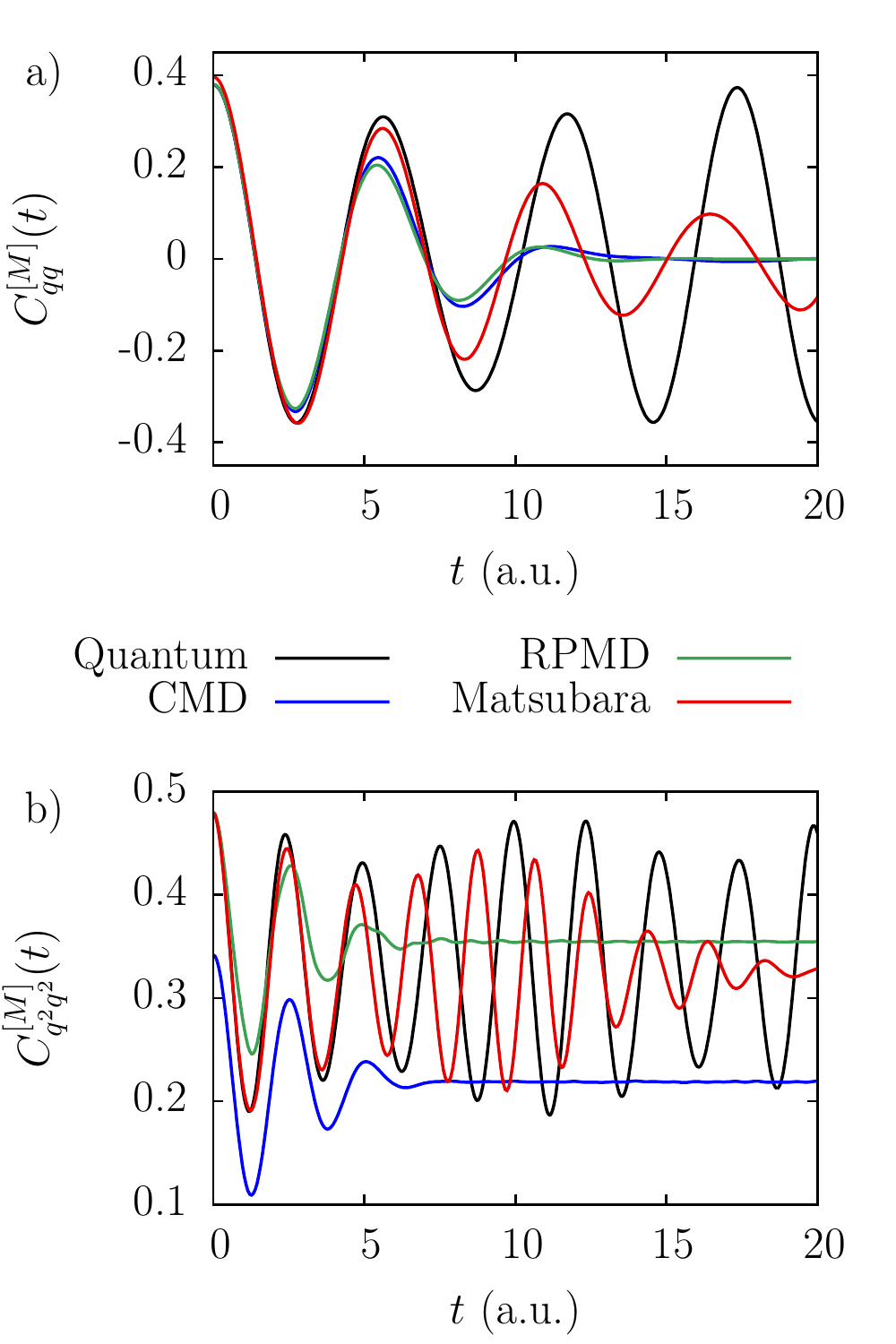}
        \caption{Comparisons of Matsubara, CMD, RPMD and  (exact) quantum
            Kubo-transformed autocorrelation functions, for the quartic
            potential $V(q)=q^4/4$, with mass $m=1$, at temperature $\beta=2$
            (in atomic units). The position autocorrelation functions in (a) are
            taken from ref.~\onlinecite{mats}. The position-squared
            autocorrelation functions in (b) were calculated numerically using
            the procedure described in ref.~\onlinecite{mats}, with $M=7$
            Matsubara modes.\cite{nonerg}  The differences between the Matsubara
            and exact quantum results show the importance of real-time quantum coherence
            in this model system, the neglect of which (in the Matsubara
            calculations) has blue-shifted and broadened the spectrum.}
        \label{figure1}
    \end{center}
\end{figure}
In the usual way of mean-field dynamics,\cite{coarse} we can split the force on the centroid into
\begin{align}
-{\partial {\widetilde U}_M({\bf \widetilde Q})\over\partial {\widetilde Q}_0}=F_0(\wt Q_0)+F_{\rm fluct}({\bf \wt Q})\label{fluct}
\end{align}
where $F_0(\wt Q_0)$ is the mean-field force 
%\begin{align}
%F_0(\wt Q_0)=-{1\over Z(Q_0)}\int d{\bf \widetilde P}'& \int d{\bf \widetilde Q}'\ e^{-\beta[{\wt H}_M({\bf \wt P},{\bf \wt Q})-i\theta_M({\bf \widetilde P},{\bf \widetilde Q})]}\,{\partial {\widetilde U}_M({\bf \widetilde Q})\over\partial {\widetilde Q}_0}\no\\
%=-{1\over Z(Q_0)}\int d{\bf \overline P}'& \int d{\bf \widetilde Q}'\ e^{-\beta \wt R_M({\bf \overline P},{\bf \wt Q})}\,{\partial {\widetilde U}_M({\bf \widetilde Q})\over\partial {\widetilde Q}_0}
%\end{align}
\begin{align}
F_0(\wt Q_0)=& -{1\over Z(Q_0)}\int d{\bf \widetilde P}' \int d{\bf \widetilde Q}'\ \no \\
             & \times e^{-\beta[{\wt H}_M({\bf \wt P},{\bf \wt Q})-i\theta_M({\bf \widetilde P},{\bf \widetilde Q})]}
              {\partial {\widetilde U}_M({\bf \widetilde Q})\over\partial {\widetilde Q}_0}\no\\
            =& -{1\over Z(Q_0)}\int d{\bf \overline P}' \int d{\bf \widetilde Q}'\ \no \\
             & \times e^{-\beta \wt R_M({\bf \overline P},{\bf \wt Q})}\,{\partial {\widetilde U}_M({\bf \widetilde Q})\over\partial {\widetilde Q}_0}
\end{align}
(and we have used the $t=0$ contour-integration trick to get to the second line), 
\begin{align}
Z(Q_0)=\int d{\bf \overline P}'& \int d{\bf \widetilde Q}'\ e^{-\beta \wt R_M({\bf \overline P},{\bf \wt Q})}
\end{align}
and  $F_{\rm fluct}({\bf \wt Q})$ is the fluctuation force (defined by \eqn{fluct} as the difference between the exact force and $F_0(\wt Q_0)$).
\Eqn{dotb} then splits into
%\begin{align}
%\dot b(\wt Q_0,\wt P_0,t)=&\left[{{\widetilde P}_0\over m}{\partial \over \partial {\widetilde Q}_0}+F_0(\wt Q_0){\partial \over \partial {\widetilde P}_0}\right]b(\wt Q_0,\wt P_0,t)\no\\
%+&\int d{\bf \widetilde P}' \int d{\bf \widetilde Q}'\ e^{-\beta[{\wt H}_M({\bf \wt P},{\bf \wt Q})-i\theta_M({\bf \widetilde P},{\bf \widetilde Q})]} 
%F_{\rm fluct}({\bf \wt Q}){\partial \over \partial {\widetilde P}_0} e^{{\cal L}_{M}t}B({\bf \widetilde Q})\label{mfluc}
%\end{align}
\begin{align}
\dot b(\wt Q_0,\wt P_0,t)=&\left[{{\widetilde P}_0\over m}{\partial \over \partial {\widetilde Q}_0}+F_0(\wt Q_0){\partial \over \partial {\widetilde P}_0}\right]b(\wt Q_0,\wt P_0,t)\no\\
&+\int d{\bf \widetilde P}' \int d{\bf \widetilde Q}'\ e^{-\beta[{\wt H}_M({\bf \wt P},{\bf \wt Q})-i\theta_M({\bf \widetilde P},{\bf \widetilde Q})]} \no \\
&\times F_{\rm fluct}({\bf \wt Q}){\partial \over \partial {\widetilde P}_0} e^{{\cal L}_{M}t}B({\bf \widetilde Q})\label{mfluc}
\end{align}
This type of expression is encountered in coarse-graining, where the integral is sometimes approximated by a generalized Langevin term.\cite{coarse} It is an exact rewriting
 of \eqn{dotb}. Neglect of the integral term gives the mean-field approximation
\begin{align}
\dot b(\wt Q_0,\wt P_0,t)\simeq&\left[{{\widetilde P}_0\over m}{\partial \over \partial {\widetilde Q}_0}+F_0(\wt Q_0){\partial \over \partial {\widetilde P}_0}\right]b(\wt Q_0,\wt P_0,t)
\end{align}
which is CMD.\cite{cmd1,cmd2,geva}

Thus CMD corresponds to approximating Matsubara dynamics by leaving out the fluctuation term in \eqn{mfluc}. 
This result is not a surprise, and is consistent with previous numerical findings\cite{marx} that CMD causes errors through neglect of fluctuations (see Sec.~V).
 What is new is that \eqn{mfluc} gives an explicit formula for these fluctuations, in the case that the quantum dynamics can be approximated by Matsubara dynamics.

\section{Analytic continuation at $t>0$}

We now return to \eqn{trans}, which is just \eqn{mattsc} rewritten in terms of $({\bf \overline P},{\bf \wt Q})$.  Expressing ${\cal L}_{M}$ in terms of these coordinates gives
\begin{align}
{\cal L}_{M}={\cal L}_M^{[\rm RP]}+i{\cal L}_M^{[\rm I]}\label{liouliou}
\end{align}
where
%\begin{align}
%{\cal L}_M^{[\rm RP]}=\sum_{n=-(M-1)/2}^{(M-1)/2}{{\overline P}_n\over m}{\partial \over \partial {\widetilde Q}_n}
%  - \left[m\wt \omega_n^2\wt Q_n+{\partial {\widetilde U}_M({\bf \widetilde Q})\over\partial {\widetilde Q}_n}\right]{\partial \over\partial {\overline P}_n}
%\end{align}
\begin{align}
{\cal L}_M^{[\rm RP]}= \!\!\!\! \sum_{n=-(M-1)/2}^{(M-1)/2} \! {{\overline P}_n\over m}{\partial \over \partial {\widetilde Q}_n} 
\! - \! \left[m\wt \omega_n^2\wt Q_n \! + \!{\partial {\widetilde U}_M({\bf \widetilde
Q})\over\partial {\widetilde Q}_n}\right]\!{\partial \over\partial {\overline P}_n}
\end{align}
is the RPMD Liouvillian (corresponding to the ring-polymer hamiltonian  $\wt R_M({\bf \overline P},{\bf \wt Q})$)  and
\begin{align}
{\cal L}_M^{[\rm I]}=\sum_{n=-(M-1)/2}^{(M-1)/2} \wt\omega_n\left(\overline P_n{\partial \over\partial {\overline P}_{-n}}-\wt Q_{n}{\partial \over\partial {\wt Q}_{-n}}\right)
\end{align}
Note that the complete Liouvillian ${\cal L}_{M}$ does not correspond to a Hamiltonian in $({\bf \overline P},{\bf \wt Q})$
(because the transformation from $({\bf \wt P},{\bf \wt Q})$ to $({\bf
\overline P},{\bf \wt Q})$ is non-canonical), and that any resemblance to
RPMD\cite{rpmd1,rpmd2} is at this stage illusory, since the imaginary parts of
${\overline P}_n$, $n\ne 0$, contribute terms that cancel the spring terms in ${\cal L}_M^{[\rm RP]}$.

\begin{table*}[t]
\renewcommand{\arraystretch}{1.7}
\small
%\hskip-2cm
\begin{tabular}{ L{8.5cm} c|c L{8.5cm}}\hline
\multicolumn{1}{c}{\bf CMD} & \ &  \ &\multicolumn{1}{c}{\bf RPMD} \\ \hline
satisfies detailed balance, because the centroid mean-field force is decoupled from the Matsubara fluctuations &&& satisfies detailed balance,
 because ${\cal L}_M^{[\rm RP]}$ and ${\cal L}_M^{[\rm I]}$ [in \eqn{liouliou}] independently satisfy detailed balance\\ 
is the centroid mean-field approximation to Matsubara dynamics &&&
has the same centroid mean-field approximation as Matsubara dynamics, namely CMD \\ 
is exact for linear TCFs in the harmonic limit, since the centroid mean-field force is then equal to the Matsubara force &&&
is exact for linear TCFs in the harmonic limit, since the neglected term ${\cal L}_M^{[\rm I]}$ does not act on the centroid\\
gives the exact centroid-averaged Matsubara Liouvillian dynamics at $t=0$ &&&
gives the exact Matsubara Liouvillian dynamics at $t=0$\\ 
%reproduces the Matsubara position ATCF up to order  $t^4$ at $t=0$, and the momentum ATCF up to order $t^2$
%(see Appendix D)&&&
%reproduces the Matsubara position ATCF up to order $t^x$ at $t=0$, the momentum ATCF to order $t^x$, the position-squared ATCF to order $x$, and a general ATCF to order $tx$ (see Appendix D) \\
suffers from  the {\em curvature problem} in vibrational spectra because of the neglect of the Matsubara fluctuations &&&
suffers from {\em spurious resonances} in vibrational spectra because the neglect of ${\cal L}_M^{[\rm I]}$ shifts the fluctuation frequencies\\
gives the mean-field-averaged Matsubara force on the centroid&&&
gives the exact Matsubara force on the centroid\\ 
breaks down completely for non-linear $\hat A$ and $\hat B$  (see Fig.~1b) because $A({\bf \wt Q})$ and $B({\bf \wt Q})$ depend on non-centroid modes&&&
breaks down more rapidly for non-linear (than for linear) $\hat A$ and $\hat B$ (see Fig.~1b)
 because the neglected term ${\cal L}_M^{[\rm I]}$ acts directly on the non-centroid modes\\\hline
\end{tabular}
\caption{Properties of CMD and RPMD derived from Matsubara dynamics (TCF = time-correlation function).}
\end{table*}

If we now try to shift  ${\overline P}_n$, $n\ne 0$, onto the real axis, we find that the dynamics generated by ${\cal L}_{M}$ propagates an initial distribution of real phase-space points into the complex plane, along unstable trajectories. We do not know whether the contour-integration trick remains valid for such trajectories; even if it does, they appear to be at least as difficult to treat numerically as the sign problem in \eqn{mattsc}.

 However,  it is possible\cite{suppl} to follow a path along which one gradually
 moves ${\overline P}_n$, $n\ne 0$, towards the real axis whilst gradually discarding 
 ${\cal L}_M^{[\rm I]}$, such that the dynamics remains stable (and the contour-integration trick remains valid) at every point along the path.
  At the end of the path, ${\cal L}_M^{[\rm I]}$ has been completely discarded,
  and ${\overline P}_n$ has reached the real axis. This results in the approximation, 
% \begin{align}
% C_{ AB}^{[M]}(t) \simeq{\alpha_M\over 2\pi\hbar}\int d{\bf \overline P}& \int d{\bf \widetilde Q}\ A({\bf \widetilde Q})e^{-\beta \wt R_M
% ({\bf \overline P},{\bf \wt Q})}e^{{\cal L}_M^{[RP]}t}B({\bf \widetilde Q})
%\end{align}
 \begin{align}
 C_{ AB}^{[M]}(t) \simeq{\alpha_M\over 2\pi\hbar}\int \!\! d{\bf \overline P}& \int
 \!\! d{\bf \widetilde Q}\ A({\bf \widetilde Q})e^{-\beta \wt R_M
 ({\bf \overline P},{\bf \wt Q})}e^{{\cal L}_M^{[\rm RP]}t}B({\bf \widetilde Q})
\end{align}
which is  RPMD.\cite{rpmd1,rpmd2,smooth2}

A harmonic analysis\cite{suppl} shows that the main effect of discarding
${\cal L}_M^{[\rm I]}$ is erroneously to shift the Matsubara fluctuation frequencies to the ring-polymer frequencies. Since ${\cal L}_M^{[\rm I]}$ does not act directly on $\wt Q_0$, it follows that an RPMD time-correlation function involving linear operators (for which  $B({\bf \wt Q}) = \wt Q_0$ or $\wt P_0$) will  agree initially with the Matsubara result, but 
will then lose accuracy as the errors in the fluctuation dynamics couple to the centroid through the
 anharmonicity in ${\widetilde U}_M({\bf \widetilde Q})$.  This result is not a surprise, as the ring-polymer frequencies are known to interfere with the centroid dynamics.\cite{marx,rpmd2} What is new is that we have identified the approximation made by RPMD, namely the neglect of ${\cal L}_M^{[\rm I]}$.

\section{Discussion}

We have shown that both CMD and RPMD are approximations to Matsubara dynamics, which, as mentioned in the Introduction, is probably the correct way to describe quantum statistics and classical dynamics.
 CMD neglects the Matsubara fluctuation term; RPMD neglects part of the Matsubara Liouvillian. So far as we can tell, there is no direct physical justification that can be given for either of these approximations. CMD and RPMD are useful because, as has long been known,\cite{cmd1,cmd2,rpmd1,rpmd2} they preserve detailed balance, and satisfy a number of important limits. These properties\cite{ramirez} (and a few others) can be rederived from Matsubara dynamics, and are listed in Table I. Note also that CMD and RPMD give the same $t=0$ leading-order error terms when compared with Matsubara dynamics as with the exact quantum dynamics. \cite{braams,propa}

One new finding, less drastic than it first appears, is that both CMD and RPMD give qualitatively wrong fluctuation dynamics at barriers. In Matsubara dynamics, some of the  distributions in $q(\tau)$ stretch indefinitely over the barrier top, such that a proportion of the distribution ends up on either side. In CMD and RPMD, all of the distribution ends up on one side of the barrier (because CMD decouples the fluctuation modes necessary for stretching over the barrier, and RPMD shifts the frequencies of these modes from imaginary to real\cite{inst}).
 However, CMD and RPMD are still powerful tools for estimating quantum reaction rates, as the exact $t=0$ behaviour of these methods (see Table I) ensures that  classical rate theory (in the mean-field centroid or ring-polymer space) gives lower bound estimates of the $t\to0_+$ quantum transition-state theory rate,\cite{tim1}  for the special case of a centroid dividing-surface (CMD), and for the general case (RPMD).

The main new result of this work is that, in relating CMD and RPMD to Matsubara dynamics, we have obtained explicit formulae for the terms that are left out, which may lead to improvements in these methods. For example, it might be possible to include approximately the Matsubara fluctuation term of
\eqn{mfluc} which is missing in CMD, or to exploit the property that RPMD gives the correct Matsubara force on the centroid.\cite{tperhaps} 

\begin{acknowledgments}
TJHH, MJW and SCA acknowledge funding from the U.K. Engineering and Physical
Sciences Research Council. AM acknowledges the European Lifelong Learning
Programme (LLP) for an Erasmus student placement scholarship. TJHH also
acknowledges a Research Fellowship from Jesus College, Cambridge and helpful
discussions with Dr Adam Harper.
\end{acknowledgments}

\appendix

  \section*{Appendix: Matsubara modes}
  
  The set of $M$ Matsubara modes ${\bf \wt Q}$ is defined as\cite{mats}
  \begin{align}
{\wt Q}_n&=\lim_{N\to\infty}{1\over\sqrt{N}}\sum_{l=1}^NT_{ln}q_l, \quad n=0,\pm 1,\dots,\pm (M-1)/2
\end{align}
where  $M$ is odd\cite{mats} and satisfies $M\ll N$;  ${\bf
q}\equiv\{q_l\},l=1,\dots, N$, are a set of discrete path-integral coordinates distributed at equally spaced intervals $\beta\hbar/N$ of imaginary time, and
%\begin{align}
% T_{ln} = 
% \left\{
% \begin{array}{ll}
%  N^{-1/2} & n=0 \\
%  \sqrt{2/N} \sin(2\pi ln/N) & n=1,\dots,(M-1)/2 \\
%  \sqrt{2/N} \cos(2\pi ln/N) & n=-1,\dots,-(M-1)/2
% \end{array}
% \right.\label{ttt}
%\end{align}
\begin{align}
\!\!\! T_{ln} \! = \!\!
 \left\{
 \begin{array}{ll}
  \!\!\! N^{-1/2} & \! n\!=\!0 \\
  \!\!\! \sqrt{2/N} \sin(2\pi ln/N) & \! n\!=\!1,\dots,(M\!-\!1)/2 \\
  \!\!\! \sqrt{2/N} \cos(2\pi ln/N) & \! n\!=\!-\!1,\dots,-\!(M\!-\!1)/2
 \end{array}
 \right.\label{ttt}
\end{align}
The momentum coordinates ${\bf \wt P}$ are similarly defined in terms of ${\bf p}$.   $\wt Q_0$ and $\wt P_0$ are the position and momentum centroid coordinates. We define the associated Matsubara frequencies $\wt \omega_n=2n\pi/\beta\hbar$ such that they include the sign of $n$, which gives $\theta_M({\bf \widetilde P},{\bf \widetilde Q})$ the simple form of \eqn{thet}.
  
The functions $A({\bf \wt Q})$ and $B({\bf \wt Q})$ in \eqn{mattsc} are obtained by making the substitutions
    \begin{align}
  q_l=\sqrt{N}\!\!\sum_{n=-(M-1)/2}^{(M-1)/2}T_{ln}{\wt Q}_n\label{qsub}
   \end{align}
into the functions
 \begin{align}
 A(\bq) =& \frac{1}{N}\sum_{l=1}^N A({q}_l),\qquad B(\bq) = \frac{1}{N} \sum_{l=1}^N B({q}_l)\label{aaa}
  \end{align}
  The Matsubara potential ${\widetilde U}_M({\bf \widetilde Q})$ is obtained similarly,
  by subsituting for $q_l$ in the ring-polymer potential
   \begin{align}
   U_N({\bf q})={1\over N}\sum_{l=1}^N V(q_l)
    \end{align}
 We emphasise that the formulae above and in Sec.~II result from just one approximation, namely decoupling
  the Matsubara modes  from the non-Matsubara modes in the exact quantum Liouvillian (which causes all Liouvillian terms
  ${\cal O}(\hbar^2)$ to vanish).\cite{mats}

\clearpage


\section*{Supplementary material}

# Supplemental material for: "Relation of centroid molecular dynamics and ring-polymer molecular dynamics to exact quantum dynamics"

Timothy J. H. Hele, Michael J. Willatt, Andrea Muolo and Stuart C. Althorpe

*Department of Chemistry, University of Cambridge, Lensfield Road, Cambridge, CB2 1EW, UK.*

## I. ANALYTIC CONTINUATION

To move $\overline{P}_n$ to the real axis at $t = 0$, we apply the standard contour-integration trick used to take the Fourier tranform of a Gaussian: We construct a rectangular contour with top and bottom sides running along $\operatorname{Im} \overline{P}_n = -m\widetilde{\omega}_n \widetilde{Q}_{-n}$ and the real axis, and vertical sections at $\operatorname{Re} \overline{P}_n = \pm L$, then take the limit $L \to \infty$. The vertical sections give zero, and there are no poles inside the contour, so the integral is unchanged on shifting $\overline{P}_n$ to the real axis.

We do not know if this trick is valid at $t > 0$, since the dynamics propagates $\overline{\mathbf{P}}$ and $\widetilde{\mathbf{Q}}$ into the complex plane, resulting in unstable trajectories, which may cause $e^{\mathcal{L}_M t} B(\widetilde{\mathbf{Q}})$ to diverge.[1] However, the dynamics does remain stable, and the analytic continuation valid, if one gradually discards $\mathcal{L}_M^{[I]}$ as one pushes $\overline{P}_n$ towards the real axis. One can do this by letting $\lambda$ change smoothly from 1 to 0 in the Liouvillian

$$\mathcal{L}_\lambda = \mathcal{L}_M^{[\mathrm{RP}]} + i\lambda \mathcal{L}_M^{[\mathrm{I}]} \tag{S1}$$

whilst setting $\overline{P}_n = \Pi_n - i\lambda m\widetilde{\omega}_n \widetilde{Q}_{-n}$ (where $\Pi_n$ and $Q_{-n}$ are real). Writing $\mathcal{L}_\lambda$ in terms of $\Pi_n$ and $Q_{-n}$, we obtain

$$\mathcal{L}_\lambda = \sum_{n=-(M-1)/2}^{(M-1)/2} \frac{\Pi_n}{m} \frac{\partial}{\partial \widetilde{Q}_n} - \left[ \frac{\partial \widetilde{U}_M(\widetilde{\mathbf{Q}})}{\partial \widetilde{Q}_n} + m(1-\lambda^2)\widetilde{\omega}_n^2 \widetilde{Q}_n \right] \frac{\partial}{\partial \Pi_n} \tag{S2}$$

which shows that the dynamics of $\overline{\mathbf{P}}$ and $\widetilde{\mathbf{Q}}$ maps onto a real dynamics in $\mathbf{\Pi}$ and $\widetilde{\mathbf{Q}}$ at every value of $\lambda$ between 1 and 0, and thus avoids the unstable trajectories in the complex plane. Note that the dynamics generated by $\mathcal{L}_\lambda$ satisfies detailed balance (because $\mathcal{L}_M^{[\mathrm{RP}]}$ and $\mathcal{L}_M^{[\mathrm{I}]}$ independently satisfy detailed balance).

## II. ANALYSIS OF HARMONIC FLUCTUATION TERMS

If the potential $V(q)$ is harmonic, then $\overline{P}_n$ in Eq. (7) of the article can be analytically continued to the real axis for $t > 0$, giving the Matsubara equations of motion

$$\widetilde{Q}_n(t) = \widetilde{Q}_n \cos \omega t + \frac{\overline{P}_n}{m\omega} \sin \omega t + i\frac{\widetilde{\omega}_n}{\omega}\widetilde{Q}_{-n} \sin \omega t$$

$$\overline{P}_n(t) = \overline{P}_n \cos \omega t - \frac{m\overline{\omega}_n^2}{\omega}\widetilde{Q}_n \sin \omega t - i\frac{\widetilde{\omega}_n}{\omega}\overline{P}_{-n} \sin \omega t \tag{S3}$$

where $\overline{\omega}_n = \sqrt{\widetilde{\omega}_n^2 + \omega^2}$ are the ring-polymer frequencies. On discarding $\mathcal{L}_M^{[I]}$, we obtain the RPMD equations of motion

$$\widetilde{Q}_n(t) = \widetilde{Q}_n \cos \overline{\omega}_n t + \frac{\overline{P}_n}{m\overline{\omega}_n} \sin \overline{\omega}_n t$$

$$\overline{P}_n(t) = \overline{P}_n \sin \overline{\omega}_n t - m\overline{\omega}_n \widetilde{Q}_n \sin \overline{\omega}_n t \tag{S4}$$

Hence the omission of $\mathcal{L}_M^{[I]}$ changes the fluctuation frequencies $\omega$ to the ring-polymer frequencies $\overline{\omega}_n$. Curiously, the amplitude of the fluctuations is unchanged, since for both Matsubara and RPMD dynamics, the $(M-1)/2$ quantities,

$$m^2 \overline{\omega}_n^2 \left( \widetilde{Q}_n^2 + \widetilde{Q}_{-n}^2 \right) + \overline{P}_n^2 + \overline{P}_{-n}^2 \tag{S5}$$

are constants of the motion. This analysis can be repeated for a parabolic barrier, where it is found that discarding $\mathcal{L}_M^{[I]}$ changes the fluctuation frequencies from the imaginary barrier frequency $i|\omega|$ to the real ring-polymer frequencies $\overline{\omega}_n = \sqrt{\widetilde{\omega}_n^2 - |\omega|^2}$ (if $\beta\hbar < 2\pi/|\omega|$).

### REFERENCES

[1] C.M. Bender, D.C. Brody and D.W. Hook, J. Phys. A: Math. Theor. **41**, 352003 (2008).




\begin{thebibliography}{10}
\bibitem{noz} To reduce clutter, we define  $C_{AB}(t)$ without the factor of $1/Z$ (where $Z$ is the quantum partition function).
\bibitem{alg}  To simplify the algebra, we assume that the system is one-dimensional, and that ${\hat A}$ and ${\hat B}$ are functions  of $\hat q$. Matsubara dynamics generalizes straightforwardly to multi-dimensions and to operators that are functions of $\hat p$; see ref.~\onlinecite{mats}.
\bibitem{smooth} Note that this distribution corresponds to the Matsubara (i.e.~smoothed) version of the 
 ring-polymer distribution of ref.~\onlinecite{freeman}
  which converges in the large $M$ limit to the same distribution as the more commonly used discrete  form of ref.~\onlinecite{chanwol,parri}.
\bibitem{geva} See also: Q.\ Shi and E.~Geva, \jcp {\bf 118}, 8173 (2003), which obtains CMD by decoupling the centroid 
 from LSC-IVR dynamics (which is equivalent to filtering out the non-Matsubara
 modes from LSC-IVR, then following the steps above).
\bibitem{smooth2} Strictly speaking, this is the Matsubara-smoothed version of RPMD, but it converges in the large $M$ limit
 to the more familiar (and practical) discrete version of RPMD in refs.~\onlinecite{rpmd1,rpmd2}; see also ref.~\onlinecite{smooth}.
\bibitem{nonerg} The RPMD, Matsubara and exact quantum results would tend to the same $t\to\infty$ limit in the case of an ergodic system (which this is not).
\end{thebibliography}
\end{document}